\documentclass{llncs}

\usepackage{jason}
\usepackage{paralist,mathtools,multirow}
\usepackage{url}
\usepackage{stmaryrd}
\usepackage{tikz}
\usepackage{arydshln,array}
\usepackage{etoolbox}

\usetikzlibrary{arrows,automata,shapes}
\usetikzlibrary{backgrounds,fit,decorations.pathreplacing}
\usetikzlibrary{positioning,arrows,shapes,fit,chains,decorations.pathmorphing,decorations.pathreplacing}

\renewcommand{\true}{1}
\renewcommand{\false}{0}

\newcommand{\allow}{\true}
\newcommand{\deny}{\false}
\newcommand{\na}{\bot}

\newcommand{\pand}{\mathbin{\land}}
\newcommand{\pdov}{\mathbin{\vartriangle}} % deny overrides (weak Kleene conjunction)
\newcommand{\paov}{\mathbin{\triangledown}} % allow overrides (weak Kleene disjunction)
\newcommand{\por}{\mathbin{\lor}}
\newcommand{\pnot}{\mathop{\lnot}}
\newcommand{\dbd}{\mathop{\sim}}

\newcommand{\pdecisions}{\mathsf{Dec}_{\rm P}}

\newcommand{\sem}[1]{\ensuremath{\llbracket #1 \rrbracket}}

\newcommand{\queries}{\ensuremath{Q}\xspace}
\newcommand{\apolicies}{\ensuremath{\mathcal{A}}\xspace}
\newcommand{\reqorder}{\ensuremath{\leqslant_\mathcal{Q}}}
\newcommand{\decorder}{\ensuremath{\leqslant}}
\newcommand{\threeorder}{\ensuremath{\leqslant_{\sf 3}}}
\newcommand{\three}{\ensuremath{\mathsf{Three}}}
\newcommand{\ptransplication}{\mathbin{?}}
\newcommand{\pinterjunction}{\mathbin{\mathrlap{\vee}{\wedge}}}

\newcommand{\conflict}{\ensuremath{\top}}
\newcommand{\four}{\ensuremath{\mathsf{Four}}}
\newcommand{\fourorder}{\ensuremath{\leqslant_{\sf 4}}}

\newcommand{\pfirst}{\mathbin{\triangleright}}

\renewcommand{\subfigcapskip}{6pt}
\newcommand{\xpermit}{\ensuremath{\mathsf{Permit}}\xspace}

\newcommand{\xna}{\ensuremath{\mathsf{NA}}\xspace}

\newcommand{\abac}{\ensuremath{\mathsf{ABAC}}\xspace}
\newcommand{\abacc}{\ensuremath{\abac_\mathsf{C}}\xspace}
\newcommand{\abacm}{\ensuremath{\abac_\mathsf{M}}\xspace}
\newcommand{\abacrequests}{\ensuremath{\mathcal{Q}^{\sf ?}}}
\newcommand{\abacreqorder}{\ensuremath{\leqslant_{\mathcal{Q}}}}
\newcommand{\abacmrequests}{\ensuremath{\mathcal{Q}^{\sf *}}}
\newcommand{\abacmreqorder}{\ensuremath{\leqslant_{\abacmrequests}}}

\begin{document}

\title{Towards A Generic Formal Framework \\for Access Control Systems}
\author{Jason Crampton\inst{1} \and Charles Morisset\inst{2}}
\institute{
 Royal Holloway, University of London \\ \email{Jason.Crampton@rhul.ac.uk}
 \and
 Newcastle University \\ \email{Charles.Morisset@ncl.ac.uk}
}
\maketitle

\begin{abstract}
There have been many proposals for access control models and authorization policy languages, which are used to inform the design of access control systems.
Most, if not all, of these proposals impose restrictions on the implementation of access control systems, thereby limiting the type of authorization requests that can be processed or the structure of the authorization policies that can be specified.
In this paper, we develop a formal characterization of the features of an access control model that imposes few restrictions of this nature.
Our characterization is intended to be a generic framework for access control, from which we may derive access control models and reason about the properties of those models.
In this paper, we consider the properties of monotonicity and completeness, the first being particularly important for attribute-based access control systems.
XACML, an XML-based language and architecture for attribute-based access control, is neither monotonic nor complete.
Using our framework, we define attribute-based access control models, in the style of XACML, that are, respectively, monotonic and complete.
\end{abstract}

\section{Introduction}

% \begin{itemize}
% \item There are plenty of access control models/languages in the literature.
% \item Each provides different features for a specific context.
% \item We propose a meta model here based on logical operators.
% \item We focus on two specific properties: completeness and monotonicity.
% \item Completeness is useful to have an expressive language.
% \item Monotonicity is useful against attribute-hiding attacks.
% \item We show that XACML is neither complete nor monotonic
% \item We propose two simple models, one complete, and the other monotonically complete.
% \end{itemize}

One of the fundamental security services in modern computer systems is \emph{access control}, a mechanism for constraining the interaction between (authenticated) users and protected resources.
Generally, access control is enforced by a trusted component (historically known as the \emph{reference monitor}), which typically implements two functions: an \emph{authorization enforcement function} (AEF) and an \emph{authorization decision function} (ADF).
%The AEF is responsible for mediating all interactions between users and resources, while the ADF is responsible for determining whether an interaction is authorized.
The AEF traps all attempts by a user to interact with a resource (usually known as a \emph{user request}) and transforms that request into one or more \emph{authorization queries} (also known as \emph{authorization requests}) which are forwarded to the ADF.
%The ADF decides whether those queries are authorized or not and returns an authorization decision to the AEF, which implements that decision.

Most access control systems are policy-based.
That is, an administrator specifies an authorization policy, which, in its simplest form, encodes those authorization requests that are authorized.
The ADF takes an authorization query and an authorization policy as input and returns an authorization decision.
For this reason, it is common to refer to the AEF and ADF as the \emph{policy enforcement point} (PEP) and \emph{policy decision point} (PDP), respectively; it is this terminology that we will use henceforth.
%The interested reader is referred to the XACML standard, which includes a reference architecture and workflow for access control systems, for further details~\cite[Figure 1]{xacml2.0}.

An authorization policy is merely an encoding of the access control requirements of an application using the authorization language that is understood by the PDP.  It is necessary, therefore, to make a distinction between an {\em ideal policy} and a \emph{realizable policy}: the former is an arbitrary function from requests to decisions; the latter is a function that can be evaluated by the PDP.  Given a particular policy language, there might be some ideal policies that are not realizable, which may be a limitation of the policy language in practice.  The access control system used in early versions of Unix, for example, is rather limited~\cite[\S 15.1.1]{Bishop02}.  An important consideration, therefore, when designing an access control system is the \emph{expressivity} of the policy language.

The increasing prevalence of open, distributed, computing environments means that we may not be able to rely on a centralized authentication function to identify authorized users.  This means that authorization decisions have to be made on the basis of (authenticated) user attributes (rather than user identities).  In turn, this means that the structure of authorization queries needs to be rather more flexible than that used in closed, centralized environments.  The draft XACML 3.0 standard, for example, uses a much ``looser'' query format than its predecessor XACML 2.0.  However, if we have no control over the attributes that are presented to the PDP, then a malicious user (or a user who wishes to preserve the secrecy of some attributes) may be able to generate authorization decisions that are more ``favorable'' by withholding attributes from the PEP~\cite{post2012,DBLP:conf/esorics/GriesmayerM13}.  A second important consideration, therefore, is whether authorization policies are guaranteed to be
``monotonic''
 in the sense that providing fewer attributes in an authorization query yields a less favorable outcome (from the requester's perspective).

There is an extensive literature on languages for specifying authorization policies, most approaches proposing a new language or an extension of an existing one. The proliferation of languages led Ferraiolo and Atluri to raise the question in~\cite{Ferraiolo:2008:MMA:1377836.1377860} of whether a {\em meta-model} for access control was needed and possible to achieve, hinting at XACML~\cite{XACML3} and RBAC~\cite{Ferraiolo92} as potential candidates.
In response, Barker proposed a meta-model~\cite{Ba09}, which sought to identify the key components required to specify access control policies, based on a term-rewriting evaluation.

In this paper, we do not present ``yet another language'' for access control policies, nor do we claim to have a ``unifying meta-model''.
We focus instead on reasoning about the properties of a language. Indeed, we advocate the idea that a language is just a tool for policy designers:
just as some programming languages are better suited to particular applications, it seems unlikely that there exists a single access control model (or meta-model) that is ideal in all possible contexts. On the contrary, we believe that providing the structure to formally analyse a language might be valuable to a policy designer, in order to understand the suitability of a particular language as the basis for a specific access control system.

We conclude this section by summarizing the structure and contributions of the paper.
In Sec.~\ref{sec:framework} we propose a general framework for access control, whose role is not to be used as an off-the-shelf language, but as a way to identify and reason about the key aspects of a language.
In Sec.~\ref{sec:monotonicity} we define monotonicity and completeness in the context of our framework.
% We show, in Sec.~\ref{sec:xacml}, that XACML can be encoded as an instance of our framework.
% We also show that the fragment of XACML without conditions is neither monotonic nor complete, thus suggesting that, at least from a formal perspective, XACML without conditions has some limitations.
Then in Sec.~\ref{sec:general-abac} we define two attribute-based models, respectively monotonic and complete, by building on existing results from the literature on multi-valued and partial logic.
The main body of the paper ends with discussions of related and future work.
%Results are stated without proofs; the on-line version of the paper~\cite{CrMo12} includes the proofs.
%The paper concludes with three appendices, whose purpose is to make the paper self-contained.

\newcommand{\eval}{\mathsf{Eval_\mathcal{A}}}
\newcommand{\evalplus}{\mathsf{Eval^+}}
\newcommand{\ops}{\mathsf{Ops}}
\newcommand{\op}{\mathsf{op}}
\newcommand{\evalops}{\mathsf{Eval_{\ops}}}
\renewcommand{\pdecisions}{\mathsf{Dec}}
\newcommand{\semantics}[1]{\llbracket #1 \rrbracket}

\section{A Framework for Defining Access Control Models}\label{sec:framework}

In this section we describe the various components of our framework and introduce our formal definition of access control models and policies. Broadly speaking, we provide a generic method for designing access control models and for furnishing access control policies, which are written in the context of a model, with authorization semantics. We also introduce the notion of an ideal policy, which is an abstraction of the requirements of an organization, and relate this concept to that of an access control policy.

%In its most abstract form, a policy is a high-level description of the security requirements of an organization.
%A policy has to be specified using the language and tools provided by the access control system chosen to enforce those security requirements.
%A policy is therefore defined using a specific language, that the system is able to interpret.
%It is essential, therefore, to recognize that there is a distinction between the desired policy and the policy that is enforced, purely because of the limitations of the system used to represent and enforce the desired policy.
%%We first wish to distinguish between the definition of the policy itself and its interpretation in a particular context.
%In general, a policy can be seen as formula, consisting of {\em atomic policies} joined together using {\em policy operators}.
\subsection{An Informal Overview}

From an external viewpoint, an access control mechanism is a process that constrains the interactions between users and data objects. Those interactions are modeled as access requests, with the mechanism incorporating two functions: one to determine whether a request is authorized or not and one to enforce that decision. The overall process must be total, in the sense that its behavior is defined for {\em every} possible interaction (which may include some default behavior that is triggered when the decision function is unable to return a decision). In general, designing a particular access control mechanism for a particular set of requests is the final concrete objective of any access control framework (although we are also clearly interested in expressing general properties of the framework).

We define an access control mechanism using an {\em access control policy}, together with an {\em interpretation function} which provides the authorization semantics for a policy. Intuitively, a policy is simply a syntactical object, built from {\em atomic policies} and {\em policy connectives}. The interpretation function provides the denotational semantics of the policy, by returning a function from requests to decision, thus defining the expected behavior of the PDP. Clearly, a policy can be interpreted in different ways, and an interpretation function can interpret different policies, as long as they are built from the same atomic policies and connectives.

An {\em access control model} defines an {\em access control language}, which consists of a set of atomic polices and policy connectives, and an interpretation function. In other words, an access control model specifies a set of access control policies and a unique way to interpret each of these policies. An {\em access control mechanism}, then, is an instance of an access control model if its policy belongs to the language of the model and if its interpretation function is that of the model.

\subsection{The Framework}

In order to provide a framework within which policies can be constructed, we introduce the notion of access control model, which is a tuple ${\cal M} = (Q, \mathcal{A}, \ops, \pdecisions, \semantics{\cdot})$, where $Q$ is a set of \emph{requests}, $\cal A$ a set of \emph{atomic authorization policies}, $\ops$ a set of \emph{policy connectives}, $\pdecisions$ a set of (authorization) \emph{decisions}, and, for each $A \in \cal A$, $\semantics{A}$ is a total function from $Q$ to $\pdecisions$ defining the \emph{evaluation} of policy $A$ for all requests in $Q$.

Each $k$-ary policy connective $\sf op$ in $\ops$ is identified with a function ${\sf op} : \pdecisions^k \rightarrow \pdecisions$. We construct an authorization policy $P$ using elements of $\cal A$ and $\ops$. We extend the evaluation function for atomic policies to arbitrary policies: that is, $\semantics{P} : Q \rightarrow \pdecisions$ provides a method of evaluating requests with respect to a policy $P$. We say that $\semantics{\cdot}$ defines the \emph{authorization semantics} of the model.

The syntax by which policies are defined and the extension of the authorization semantics for atomic policies to non-atomic policies are fixed (for all models), as specified in Definition~\ref{def:policy-term} below. Nevertheless, different choices for $\pdecisions$, $\mathcal{A}$ and $\semantics{\cdot}$ give rise to very different models having very different properties.
%(see Appendix~\ref{app:examples} for some examples).

A \emph{policy term} $P$ is defined by a (rooted) \emph{policy tree}, in which leaf nodes are \emph{atomic policies} and each non-leaf node is a policy connective (we may also use the term \emph{policy operator}). More formally we have the following definition:
\begin{Def}
	\label{def:policy-term} Let ${\cal M} = (Q, \mathcal{A}, \ops, \pdecisions, \semantics{\cdot})$ be a model. Then every atomic policy in $\cal A$ is a \emph{policy term}. If $P_1,\dots,P_k$ are policy terms, then for each $k$-ary operator ${\sf op} \in \ops$, ${\sf op}(P_1,\dots,P_k)$ is a policy term. For each policy term ${\sf op}(P_1,\dots,P_k)$, we define
	\begin{equation}
		\label{eq:policy-semantics} \semantics{{\sf op}(P_1,\dots,P_k)}(q) = {\sf op}(\semantics{P_1}(q),\dots,\semantics{P_k}(q)).
	\end{equation}
\end{Def}

In other words, authorization policies are represented as policy trees and policies are evaluated from the bottom up by
\begin{inparaenum}
	[(a)]
	\item evaluating atomic policies
	\item combining the decisions returned for atomic policies using the relevant policy connectives.\footnote{Strictly speaking, we should use different symbols for a policy connective and the decision operator with which it is associated. We have chosen to abuse notation in the interests of clarity, because little is gained by strict adherence to formality here.}
\end{inparaenum}
We write $\mathcal{P}(\mathcal{M})$ to denote the set of policies that can be expressed within $\cal M$.

Given a set of queries $Q$ and a set of decisions $\pdecisions$, an \emph{ideal access control policy} is a total function $\pi : Q \to \pdecisions$.%
\footnote{Clearly, a policy designer could define a policy extensionally, simply by associating each query with a decision.
	  However, in practice, policies are constructed in a modular fashion, where each component defines a particular security concern and the decisions from different components are combined.}
We say that an ideal policy $\pi$ is {\em realizable} by an access control model ${\cal M}$ if, and only if, there exists a policy term $P \in \mathcal{P}(\mathcal{M})$ such that for any query $q$, $\pi(q) = \semantics{P}(q)$; in the interests of simplicity we will abuse notation and write $\pi \in \mathcal{P}(\cal M)$ and $\pi = ({\cal M}, P)$.

% A policy decision point (PDP) must implement policy evaluation using policy tree traversal and the rule in~\eqref{eq:policy-semantics}. In addition, the PDP must implement some method for computing $\semantics{A}$ given a query $q$. This second aspect of the PDP's operation is application-dependent and will be determined by the way in which queries and atomic requests are represented and compared.

%
%Figure~\ref{fig:three-operators} shows one unary and two binary operators defined over a three-valued decision set comprising the decisions $\allow$, $\deny$ and $\na$, which may be read as ``allowed'', ``denied'' and ``neither-allowed-nor-denied''.
%We will see in due course that all three operators can be useful in practice.
%A policy that returns $\na$ for request $q$ is usually said to be ``not-applicable''.
%
%\begin{figure}[h] \centering
%  \[
%    \begin{array}{c|c}
%      x & \dbd x \\
%    \hline
%      0 & 0 \\
%      1 & 1 \\
%      \bot & 0 \\
%    \end{array}
%  \qquad
%    \begin{array}{c|ccc}
%      \pdov & 0 & 1 & \bot \\
%    \hline
%      0 & 0 & 0 & 0 \\
%      1 & 0 & 1 & 1 \\
%      \bot & 0 & 1 & \bot \\
%    \end{array}
%  \qquad
%    \begin{array}{c|ccc}
%      \pand & 0 & 1 & \bot \\
%    \hline
%      0 & 0 & 0 & 0 \\
%      1 & 0 & 1 & \bot \\
%      \bot & 0 & \bot & \bot \\
%    \end{array}
%  \]
%\caption{Three logical operators on $\set{0,1,\bot}$} \label{fig:three-operators}
%\end{figure}

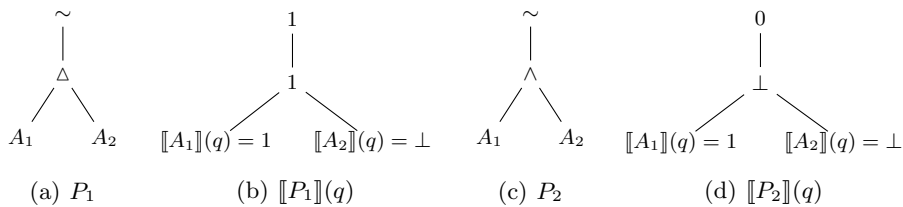
\begin{figure}
	[h] \centering
	\begin{minipage}
		{.15
		\textwidth}\centering \subfigure[$P_1$]{
		\begin{tikzpicture}[node distance=0.5cm and 0.075cm,scale=.9,transform shape]
			\node (a1) {$A_1$};
			\node[above right=of a1] (and) {$\pdov$};
			\node[below right=of and] (a2) {$A_2$};
			\node[above=of and] (dbd) {$\dbd$};
			\draw (a1) to (and);
			\draw (a2) to (and);
			\draw (and) to (dbd);
		\end{tikzpicture}
		}
	\end{minipage}
	\hfill
	\begin{minipage}
		{.32
		\textwidth}\centering \subfigure[$\semantics{P_1}(q)$]{\label{subfig:pol-tree-eval1}
		\begin{tikzpicture}[node distance=0.5cm and 0.1cm,scale=.9,transform shape]
		  \node[inner sep=0pt] (a1) {$\semantics{A_1}(q) = 1$};
		  \node[above right=of a1] (and) {$1$};
		  \node[below right=of and,inner sep=0pt] (a2) {$\semantics{A_2}(q) = \bot$};
		  \node[above=of and] (dbd) {$1$};
		  \draw (a1) to (and);
		  \draw (a2) to (and);
		  \draw (and) to (dbd);
		\end{tikzpicture}
		}
	\end{minipage}
	\hfill
	\begin{minipage}
		{.15
		\textwidth}\centering \subfigure[$P_2$]{
		\begin{tikzpicture}[node distance=0.5cm and 0.075cm,scale=.9,transform shape]
		  \node (a1) {$A_1$};
		  \node[above right=of a1] (and) {$\pand$};
		  \node[below right=of and] (a2) {$A_2$};
		  \node[above=of and] (dbd) {$\dbd$};
		  \draw (a1) to (and);
		  \draw (a2) to (and);
		  \draw (and) to (dbd);
		\end{tikzpicture}
		}
	\end{minipage}
	\hfill
	\begin{minipage}
		{.32
		\textwidth}\centering \subfigure[$\semantics{P_2}(q)$]{\label{subfig:pol-tree-eval2}
		\begin{tikzpicture}
			[node distance=0.5cm and 0.1cm,scale=.9,transform shape] \node[inner sep=0pt] (a1) {$\semantics{A_1}(q) = 1$}; \node[above right=of a1] (and) {$\bot$}; \node[below right=of and,inner sep=0pt] (a2) {$\semantics{A_2}(q) = \bot$}; \node[above=of and] (dbd) {$0$}; \draw (a1) to (and); \draw (a2) to (and); \draw (and) to (dbd);
		\end{tikzpicture}
		}
	\end{minipage}
	\caption{Illustrative policy trees and their evaluation} \label{fig:pol-tree}
\end{figure}

Figure~\ref{fig:pol-tree} shows two policy trees each having the same atomic policies, $A_1$ and $A_2$.
The figure also shows two evaluations of the tree for the same request $q$, where $\semantics{A}_1(q) = \allow$ and $\semantics{A_2}(q) = \na$. The symbols $1$, $0$ and $\na$ denote allow, deny and inapplicable decisions, respectively. The policy trees are evaluated using a post-order traversal, in which each leaf node is assigned a value according to the semantics defined by $\semantics{\cdot}$ and each interior node is assigned a value by combining the values assigned to its child nodes.
The policies in Figure~\ref{fig:pol-tree} make use of three operators taken from Table~\ref{tab:operators}.
Both $\pdov$ and $\pand$ are similar to the allow-overrides operator familiar from XACML (and also the two conjunction operators from Kleene's 3-valued logic) and only differ in the way in which $\bot$ is combined with $\allow$.
The $\dbd$ unary operator implements a deny-by-default rule, thus $\semantics{P_1}(q) \ne \semantics{P_2}(q)$.

% An \emph{access control model} is an instance of the metamodel in which
%   \begin{inparaenum}[(i)]
%     \item a particular set of connectives and a particular set of decisions are chosen
%     \item languages for representing queries and atomic policies are specified
%     \item a method of realizing the partial function $\semantics{A}$ is defined.
%   \end{inparaenum}
% An \emph{access control system} is an implementation of a model in which the semantics for atomic policies and policy terms are implemented and a policy term is chosen.
% Concrete queries are evaluated with respect to the policy term.
In general, an access control model does not specify any policy in particular (unless the language is so restricted that it can only specify one policy). To some extent, an access control model (in the sense in which we use the term in this paper) is analogous to a programming language: it describes the syntax that is used to build access control policies (analogous to programs) and the semantics of the run-time mechanisms that will be used to handle input data (access control requests in this context). A realizable policy is in this case analogous to a program $P$ written in the syntax of the model ${\cal M}$, that is interpreted using the authorization semantics of the model, while an ideal policy is analogous to the set of functional requirements.

% \subsection{Ideal and Realizable Policies}

% A multi-user computer system will generally require some form of access control in order to restrict access to protected resources.
% Generally speaking, access control is achieved by defining a policy that can be interpreted by an access control mechanism.
% In the context of our framework, an access control mechanism is an implementation of the authorization semantics function for a particular model $\cal M$ and an access control policy is a member of $\mathcal{P}(\cal M)$.
% However, the model may not be sufficiently expressive to encode the access control requirements.
% The Unix access control model is an obvious example of a model that is unable to articulate certain access control requirements.
% More formally, g

Note that an ideal policy can be realized by different access control models: $\pi \in \mathcal{P}(\mathcal{M})$ and $\pi \in \mathcal{P}(\mathcal{M}')$ with $\cal M \ne \cal M'$. In other words, different access control mechanisms may be able to enforce the same security requirements. And $\pi$ may be realizable by different policy terms from the same access control model: $\pi = (\mathcal{M},P)$ and $\pi = (\mathcal{M},P')$ with $P \ne P'$. In other words, security requirements can be enforced by the same mechanism using different policies. However, an ideal policy may not be realizable by any policy term for a given model; the extent to which a model can realize the set of ideal policies provides us with a notion of the \emph{completeness} of a model (as we discuss in Section~\ref{sec:completeness}).

\subsection{Framework Instantiation}
\label{sec:examples}

\newcommand{\AM}{\ensuremath{\mathsf{AM}}} A model provides the global structure from which access control policies can be built. A simple example of a model is the protection matrix model~\cite{harr:prot76}, which can be viewed as a set of triples $(s, o, x)$, where $s$ is a subject, $o$ an object and $x$ an access mode. A query is also a triple $(s, o, x)$, and is authorized if, and only if, it belongs to the set representing the matrix. Hence, we define the set of queries $\queries_\AM$ to be the set of all triples $(s, o, x)$, the set of decisions $\pdecisions_\AM = \set{\allow, \deny}$, where $\allow$ stands for an authorized access and $\deny$ for a denied one, the set of atomic policies $\apolicies_\AM = \queries_{AM} \cup \set{\deny}$, the set of operators $\ops_\AM= \set{\lor}$, where $\lor$ is the standard boolean disjunction, and the interpretation function $\semantics{\cdot}_\AM$ to be:
\[ \semantics{p}_\AM(q) =
\begin{cases}
	\allow & \text{if $p = q$}\\
	\deny & \text{otherwise}.
\end{cases}
\]
For instance, the policy authorizing only the accesses $(s_1, o_1, x_1)$ and $(s_2, o_2, x_2)$ can be defined as $(s_1, o_1,x_1) \lor (s_2, o_2, x_2)$.

Models can also consider richer sets of queries. Indeed, recent work considers the possibility that, in order to make a decision, an access control system might require more attributes than the traditional subject-object-action triple~\cite{post2012,XACML3,Rao:2009:AFI:1542207.1542218}. In order to define requests and atomic policies it is necessary to identify sets of attributes and the values that each of those attributes may take. Role-based access control, to take a simple example, defines the sets of roles, users and permissions, together with user-role and permission-role assignment relations.

We now introduce the notions of attribute vocabulary and attribute-based access control, which are intended to be as general as possible and allow for the construction of requests and policies.

\begin{Def}\label{def:abac}
Let $\mathcal{N}$ denote a set of attribute names, and $\mathcal{D}$ denote a set of attribute domains.
Let ${\sf dom} : {\cal N} \rightarrow {\cal D}$ be a function, where $\mathsf{dom}(\alpha)$ denotes the set of attribute values associated with attribute $\alpha$.
Then $(\mathcal{N},\mathcal{D},{\sf dom})$ defines an \emph{attribute vocabulary}.
\end{Def}

When no confusion can occur, we will simply write $\cal N$ to denote an attribute vocabulary.
A request is modeled as a set of name-value pairs of the form $(\alpha,v)$, where $\alpha \in \mathcal{N}$.
We denote the set of requests by $\mathcal{Q}^*(\mathcal{N})$, omitting $\mathcal{N}$ when it is obvious from context.
We say an  attribute name-value pair $(\alpha,v)$ is \emph{well-formed} if $\alpha \in \mathcal{N}$ and $v \in \mathsf{dom}(\alpha)$.
We assume that a PDP can recognize (and discard) name-value pairs in a request that are not well-formed.

% We write $\mathcal{A}(\mathcal{N})$ to denote the set of well-formed attribute name/value pairs, i.e., $\mathcal{A}(\mathcal{N}) = \set{(\alpha,v) : \alpha \in \mathcal{N},v \in \mathsf{dom}(\alpha)}$.
% Finally, we write $\abacmrequests({\cal N}) = \wp(\mathcal{A}(\mathcal{N}))$ for the set of attribute queries, which consist of sets of pairs attribute name/values.
%
% \begin{Def}
% Given a vocabulary $(\mathcal{N},\mathcal{T},\mathcal{D},{\sf type},{\sf dom})$, a \emph{well-formed authorization query} is a set $\set{(\alpha_1,v_1),\dots,(\alpha_m,v_m)}$, where $\alpha_i$ is an attribute type and $v_i \in {\sf dom}(\alpha_i)$.
% \end{Def}
Attribute-based access control (ABAC) policies are modular.
Hence, a policy component may be incomplete or two policy components may return contradictory decisions.
Thus, it is common to see additional decisions used to denote a policy ``gap'' or ``doubt'' indicating different reasons why policy evaluation could not reach a conclusive (allow or deny) decision~\cite{DBLP:journals/tissec/BrunsH11,Rao:2009:AFI:1542207.1542218}.
We write $\three = \set{\allow,\deny,\na}$, where $\semantics{A}(q) = \na$ indicates that $\semantics{A}(q)$ is neither $0$ nor $1$.
%Rao et al.~\cite{Rao:2009:AFI:1542207.1542218}, for example, consider the set of decisions $\set{Y, N, \mathit{NA}}$, where $Y$ indicates that a request is authorized, $N$ that a request is denied and $\it NA$ that the policy is not applicable.

\begin{table}[tp] 
  \[   
  \begin{array}{>{~}c<{~}>{~}c<{~}||>{~}c<{~}|>{~}c<{~}||>{~}c<{~}>{~}c<{~}|>{~}c<{~}>{~}c<{~}|>{~}c<{~}|>{~}c<{~}|>{~}c<{~}}
    d_1 & d_2 & \lnot d_1 & \sim d_1 & d_1 \land d_2 & d_1 \pdov d_2 & d_1 \lor d_2 & d_1 \paov d_2 & d_1 \ptransplication d_2 & d_1 \pinterjunction d_2 & d_1 \pfirst d_2 \\
    \hline \allow & \allow & \deny & \allow & \allow & \allow & \allow & \allow & \allow & \allow & \allow \\
    \allow & \deny & \deny & \allow & \deny & \deny & \allow & \allow & \deny & \na & \allow \\
    \allow & \na & \deny & \allow & \na & \allow & \allow & \allow & \na & \na & \allow \\
    \deny & \allow & \allow & \deny & \deny & \deny & \allow & \allow & \na & \na & \deny \\
    \deny & \deny & \allow & \deny & \deny & \deny & \deny & \deny & \na & \deny & \deny \\
    \deny & \na & \allow & \deny & \deny & \deny & \na & \deny & \na & \na & \deny \\
    \na & \allow & \na & \deny & \na & \allow & \allow & \allow & \na & \na & \allow \\
    \na & \deny & \na & \deny & \deny & \deny & \na & \deny & \na & \na & \deny \\
    \na & \na & \na & \deny & \na & \na & \na & \na & \na & \na & \na\\
  \end{array}	 
  \]
  \caption{Operators over $\set{\allow, \deny, \na}$} \label{tab:operators}
\end{table}

In Table~\ref{tab:operators} we summarize the characteristics of some useful 3-valued operators, most of which are self-explanatory.
The $\ptransplication$ operator acts as a policy filter: $\semantics{p_1 \ptransplication p_2} = \semantics{p_2}$ if $\semantics{p_1} = \allow$, and evaluates to $\na$ otherwise.
The $\pinterjunction$ operator models policy unanimity: $p_1 \pinterjunction p_2$ evaluates to a conclusive decision only if both $p_1$ and $p_2$ do.
In Sec.~\ref{app:conflict-abac} we describe a model with a 4-valued decision set.

ABAC is designed for open distributed systems, meaning that authenticated attributes and policy components may need to be retrieved from multiple locations.
Thus, some languages assume that policy evaluation may fail: it may be, for example, that a policy server or policy information point is down.
PTaCL~\cite{post2012} relies on a three-valued logic, and considers sets of decisions in order to model indeterminacy.
XACML 3.0~\cite{XACML3} considers a six-valued decision set, three of those decisions representing different indeterminate answers.

\section{Monotonicity and Completeness}
\label{sec:monotonicity}

An access control model provides a policy designer with a language to construct a policy.
That language may well have an impact on the policies that can be expressed and the properties of those policies.
%The properties of the language might have an impact on the properties of the policies written with this language. For instance, as a trivial example, a model with a single decision ensures that only this decision can be returned by a policy.
In this section we study two specific properties of access control models, monotonicity (a kind of safety property) and completeness (an expressivity property), and we present two models satisfying these properties in Section~\ref{sec:general-abac}.

% An access control model ${\cal M} = (\queries,\apolicies,\pdecisions,\ops,\semantics{\cdot})$ may realize a large number of policies.
% It might therefore be useful to be able to characterize some general properties of the policies that belong to $\mathcal{P}(\mathcal{M})$.
% In this section, we focus on monotonicity and completeness.
% We conclude the section with a brief survey of related concepts in the literature and the connections with our work.

\subsection{Monotonicity}

Informally, a policy is monotonic whenever removing information from a request does not lead to a ``better'' policy decision. 
Such a property is of particular relevance in open systems, where users might be able to control what information they supply to the access control mechanism. A model in which all realizable policies are monotonic implies that they are not vulnerable to attribute hiding attacks~\cite{post2012}.
That is, a malicious user gains no advantage by suppressing information when making a request. 
% For instance, the policy $p_3$, defined in Section~\ref{sub:xacml_non_monotonic} in order to illustrate that XACML is not monotonic, provides an example of a policy denying explicitly a query containing a particular value for a particular attribute, and allowing all others. 
% A malicious user able to withhold that particular value (which could denote a conflict-of-interest, for instance) is therefore granted a query that should have been denied. 

We model information hiding using a partial ordering $\reqorder$ on $\queries$; the intuitive interpretation of $q \reqorder q'$ is that $q$ contains less information than $q'$.
For instance, an attribute query $q$ is less than another query $q'$ when $q \subseteq q'$.
We also need to specify what it means for a decision to ``benefit'' a user, and thus we assume the existence of an ordering relation $\decorder$ on $\pdecisions$; again, the intuitive interpretation of $d_1 \decorder d_2$ is that the decision $d_2$ is of greater benefit than $d_1$.\footnote{Note that we consider this relation to be statically defined over decisions, and to be independent of the request.}
For instance, we can consider the ordering $\threeorder$ over
$\set{\allow,\deny,\na}$, such that $x \threeorder y$ if and only if $x = y$ or $x = \na$.

\begin{Def}
Given a set of authorization queries $(\queries, \reqorder)$ and a set of decisions $(\pdecisions, \decorder)$,
a policy $\phi : \queries \to \pdecisions$ is \emph{monotonic} if, and only if, for all $q,q' \in \mathcal{Q}$, $q \reqorder q'$ implies $\phi(q) \decorder \phi(q')$.
We say that an access control model  $\mathcal{M} = (\queries,\apolicies,\pdecisions,\ops,\semantics{\cdot})$ is \emph{monotonic} if for all $P \in \mathcal{P}(\mathcal{M})$, $\semantics{P}$ is monotonic.
\end{Def}

Note that our definition of a monotonic policy applies equally well to an ideal policy $\pi : \queries \rightarrow \pdecisions$ or a realizable policy term $P$ with authorization semantics $\semantics{P} : \queries \rightarrow \pdecisions$.
%
% For instance, it is easy to see that the operators $\pnot$, $\land$ and $\lor$ defined in Table~\ref{tab:operators} are monotonic, while the operators $\dbd$, $\paov$ and $\pdov$ are not.\footnote{Recall that we are using the degree-of-definedness ordering on $\pdecisions$.}
% Using the notion of monotonic operators given in Definition~\ref{def:op-monotonicity}, we can easily prove the following proposition using induction over the structure of the policy.
% \begin{Pro}
% \label{thm:monotonicity}
% An access control model $(\queries,\apolicies,\pdecisions,\ops,\semantics{\cdot})$ is monotonic if, and only if for every atomic policy $A$ in $\mathcal{A}$, $\semantics{A}$ is monotonic and every operator in $\mathsf{Ops}$ is monotonic.
% \end{Pro}
% \begin{proof}
%   The result follows by a standard induction over the structure of policies.
% \end{proof}
%
However, the notion of monotonicity is dependent on the request ordering. For instance, without further characterization, the request ordering for the access matrix could be reduced to equality, making any policy trivially monotonic. However, more complex situations can be considered by adding extra information, such as an ordering over subjects or objects.

Tschantz and Krisnamurthi have shown that XACML 2.0 is not monotonic (although they called the property ``safety'' rather than monotonicity)~\cite{confsacmatTschantzK06}.
We show in Section~\ref{sec:abacm}---provided certain restrictions are imposed on the structure of requests---that it is possible to develop a monotonic, attributed-based (XACML-like) access control model, using results from partial logic~\cite{blamey02}.

\subsection{Completeness}\label{sec:completeness}

Given a model $\mathcal{M} = (\queries,\apolicies,\pdecisions,\ops,\semantics{\cdot})$, any realizable policy $P \in \mathcal{P}(\mathcal{M})$ clearly corresponds to an ideal policy $\pi : \queries \to \pdecisions$.
However, there may exist an ideal policy  $\pi$ (for $\queries$ and $\pdecisions$) that does not belong to $\mathcal{P}(\mathcal{M})$ and cannot, therefore, be enforced by the policy decision point.
Trivially, for example, a model without any atomic policies does not realize any policies.
It follows that the set of ideal policies that can be realized by a model represents an intuitive notion of expressivity.
A model that can realize every ideal policy is said to be complete.
More formally:
% In abstract terms, a policy is simply a function taking a request as input and returning a decision.
% Hence, an interesting question is to know whether an access control system can define any possible policy.
% More precisely, we define the notion of {\em completeness} for a system:

\begin{Def}
An access control model   $\mathcal{M} = (\queries,\apolicies,\pdecisions,\ops,\semantics{\cdot})$ is \emph{complete} if, and only if for any ideal policy $\pi : \queries \to \pdecisions$, $\pi \in \mathcal{P}(\mathcal{M})$.
\end{Def}

The completeness of a model  $(\queries,\apolicies,\pdecisions,\ops,\semantics{\cdot})$ will depend on the authorization vocabulary, the definition of atomic policies, the set $\ops$ and $\semantics{\cdot}$.
The access matrix model defined in Section~\ref{sec:examples}, for example, is complete.
  \begin{Pro}\label{thm:am_complete}
    The model $(\queries_\AM, \apolicies_\AM,  \pdecisions_\AM, \ops_\AM, \semantics{\cdot}_\AM)$ is complete.
  \end{Pro}
%   \begin{proof}
%     The proof (and all other proofs in this paper) can be found in Appendix.
%   \end{proof}
  % \begin{proof}
  %   Let $\pi : \queries_\AM \to \pdecisions_\AM$ be an ideal policy. Let $Q_{\allow}$ be the set of queries such that for any $q$ in $Q_{\allow}$, $\pi(q) = \allow$.
  %   If $Q_{\allow} = \emptyset$, then we define the policy $P = \deny$, and it is easy to see that $\semantics{P}_\AM(q) = \pi(q)$, for any request $q$.
  %   If $Q_{\allow}  = \set{q_1, \cdots, q_n}$, with $n \geqslant 1$, then we define $P = q_1 \lor \dots \lor q_n$, and we can also conclude that $\semantics{P}_{AM}(q) = \pi(q)$, for any request $q$.
  % \end{proof}

On the other hand, it is easy to show that XACML is not complete, unless we allow the inclusion of XACML conditions, which are arbitrary functions. Indeed, consider two attributes $\alpha_1$ and $\alpha_2$ with two respective attribute values $v_1$ and $v_2$, it is not possible to construct a policy that evaluates $q_1 = \set{(\alpha_1, v_1)}$ to \xpermit and $q_2 = \set{(\alpha_1, v_1), (\alpha_2, v_2)}$ to \xna, intuitively because any target not applicable to $q_2$ cannot be applicable to $q_1$. 

%A complete access control model need not be as simple as the access matrix model.
We propose an attribute-based access control model in Section~\ref{sec:abacc} in which the representation of atomic policies can distinguish attribute name-value pairs, from which we can prove a completeness result.
However, it is worth observing that, in general, if a model is both monotonic and complete, then the ordering over requests is limited to the identity relation, as illustrated above with the access matrix.

\begin{Pro}\label{thm:monotonicity-completeness}
  Given any model $\mathcal{M} = ((\queries,\reqorder),\apolicies,(\pdecisions,\decorder),\ops,\semantics{\cdot})$, if  $\mathcal{M}$ is complete and monotonic and if $\card{\pdecisions} > 1$, then $\reqorder$ is the identity relation.
\end{Pro}
% \begin{proof}
%   Let us consider two requests $q$ and $q'$ such that $q \reqorder q'$ and $q \neq q'$. Since $\card{\pdecisions} > 1$, there exist $d_1$ and $d_2$ such that $d_1 \not \decorder d_2$.
% Since $\mathcal{M}$ is complete, there exists a policy $P \in \mathcal{P}(\mathcal{M})$ such that $\semantics{P}(q) = d_1$ and $\semantics{P}(q') = d_2$, which contradicts the monotonicity of $P$.
% \end{proof}

Informally, this result states that if we wish to have a (non-trivial) monotonic model then we cannot expect to have a complete model.
Instead, what we should aim for is a model that realizes at least all \emph{monotonic ideal policies}, and such a model is said to be {\em monotonically-complete}.
In Section~\ref{sec:general-abac}, we show how to define monotonically-complete and complete attribute-based access control models that have similar characteristics to XACML.

\section{Designing Attribute-based Access Control Models}\label{sec:general-abac}

%We showed in the previous section that XACML was neither monotonic nor complete. 
It could be argued that the main objective of XACML is to provide a standard addressing as many practical concerns as possible, rather than a language with formal semantics. Nevertheless, the design choices can and should be analyzed with respect to the properties they entail. We do not claim here that XACML {\em should} be monotonic, complete, or monotonically-complete, but we show instead how, building from existing logical results, one can instantiate an access control model with these properties.

The results in this section can provide guidance to the designer of an access control system.
She can choose, for example, between a system that realizes only and all monotonic policies, and a system in which all policies are realizable, but some may be non-monotonic.
Clearly, the choice depends on the demands of the application and the constraints of the underlying environment.
While we cannot make this choice for the policy designer, our framework can only help her make an informed decision.

If the attribute vocabulary were countably infinite (and the cardinality of the decision set is greater than $1$) then the number of ideal policies would be uncountably infinite (by a standard diagonalization argument).
However, the number of realizable policies can, at best, be countably infinite, by construction.
Accordingly, it is only meaningful to consider completeness if we assume that the attribute vocabulary is finite (but unbounded).
In practice, of course, all attribute values will be stored as variables and there will be an upper limit on the size of such variables, so the attribute vocabulary will be finite and bounded, albeit by a very large number. 

\subsection{$\abacm$: A monotonic monotonically-complete model}\label{sec:abacm}

Recall from Definition~\ref{def:abac} that, given a vocabulary $\mathcal{N}$, we write $\mathcal{Q}^*(\mathcal{N})$ to denote the set of requests.
Note that a request may contain (well-formed) pairs $(\alpha,v_1,),\dots,(\alpha,v_n)$ having the same attribute name and different values.
One obvious example arises when $\alpha$ is the ``role'' attribute name and $v_i$ is the identifier of a role.
We define the set of atomic policies $\mathcal{A}(\mathcal{N})$ to be the set of well-formed name-value pairs.
That is $\mathcal{A}(\mathcal{N}) = \{(\alpha,v) : \alpha \in \mathcal{N}\ \text{and}\ v \in \mathsf{dom}(\alpha)\}$.
%The semantics of an atomic policy also use this set to define atomic policies, such that the evaluation of an atomic policy $(\alpha, v)$ given a request $q$ returns $\allow$ whenever $(\alpha, v)$ belongs to $q$, $\na$ if $q$ contains no pair of the form $(\alpha,v')$ for any $v' \in \mathsf{dom}(\alpha)$, and $\deny$ otherwise.
Then we define
\[
\semantics{(\alpha,v)}(q) =
\begin{cases}
  \allow & \text{if $q \ni (\alpha,v')$ and $v = v'$}, \\
  \na & \text{if $q \not\ni (\alpha,v')$}, \\
  \deny & \text{otherwise}.
\end{cases}
\]
Note that the above interpretation of atomic policies is by no means the only possibility.
In the context of a three-value decision set, we might return $\deny$ if $q \ni (\alpha,v')$ and $v \ne v'$, $\na$ if $q \not\ni (\alpha,v')$ and $\allow$ otherwise.
In the context of a four-value decision set, we could return $\top$ if $q \supseteq \{(\alpha,v'),(\alpha,v)\}$, since such a request both matches and does not match the attribute value $v$ for attribute $\alpha$.
We discuss these possibilities in more detail in Sec.~\ref{app:conflict-abac}.

The ordering on $\mathcal{Q}^*$, denoted by $\reqorder$, is simply subset inclusion.
We define the ordering $\threeorder$ on $\three$, where $x \threeorder y$ if and only if $x = y$ or $x = \na$. It is worth observing that if a request contains at most one value for each attribute, then each atomic policy is monotonic. More formally, if we define the set of queries $\abacrequests = \set{q \subseteq \mathcal{A}(\mathcal{N}) \mid \forall \alpha\,\, (\alpha, v) \in q \land (\alpha, v') \in q \Rightarrow v = v'}$, we can prove the following proposition. 

\begin{Pro}\label{thm:single_monotonicity}
For all requests $q, q' \in \abacrequests$ such that $q \abacreqorder q'$ and for all atomic policies $(\alpha,v) \in \mathcal{A}(\mathcal{N})$,
  we have $\semantics{(\alpha,v)}(q) \threeorder \semantics{(\alpha,v)}(q')$.
\end{Pro}

We will see in the following section that we can define a complete ABAC model that accepts requests from $\abacmrequests$, but we can no longer ensure monotonicity.
We now define a monotonic and monotonically-complete attribute-based access control (ABAC) model.

\begin{Def}
\abacm is defined to be $(\abacrequests, \mathcal{A}(\mathcal{N}), \three, \set{\pnot,\pand,\por,\pinterjunction, \ptransplication}, \semantics{\cdot})$.
\end{Def}

\abacm is not merely of academic interest because it incorporates a number of features that are similar to XACML.
In particular, we can  
  \begin{itemize}
    \item construct targets from conjunctions and disjunctions of atomic policies; % (using a method similar to that described in Section~\ref{sec:encoding-xacml});
    \item use the operators $\pand$ and $\por$ to model deny-overrides and allow-overrides policy-combining algorithms;
    \item construct (XACML) rules, policies and policy sets using policies of the form $p_1 \ptransplication p_2$, since $\semantics{p_1 \ptransplication p_2} = \semantics{p_2}$ if $\semantics{p_1} = 1$ (corresponding to ``matching'' a request to ``target'' $p_1$ and then evaluating policy $p_2$).
  \end{itemize}
The correspondence between \abacm and XACML cannot be exact, given that XACML is not monotonic.
The main difference lies in the way in which $\pand$ and $\por$ handle the $\na$ decision.
%The incompleteness of XACML is due to the way in which it deals with the $\xna$ decision.)
The operators $\pand$ and $\por$ are what Crampton and Huth called intersection operators~\cite{CH10}, whereas the policy-combining algorithms in XACML are union operators.
Informally, an intersection operator requires both operands to have conclusive decisions, while a union operator ignores inconclusive decisions.
Thus, for example, $\allow \pand \na = \na$, whereas the XACML deny-overrides algorithm would return $\allow$ given the same arguments.

A practical consequence of the design goals of \abacm is that the $\na$ decision will be returned more often than for analogous policies in XACML (or other non-monotonic languages).
In practice, the policy enforcement point will have to either 
  \begin{inparaenum}[(a)]
   \item ask the requester to supply additional attributes in the request; or
   \item deny all requests that are not explicitly allowed.
  \end{inparaenum}

\begin{Thm}\label{thm:abac-monotonically-complete}
\abacm is monotonic and monotonically complete.
\end{Thm}
\begin{proof}
Let us first observe that the operators $\pnot$,$\pand$,$\por$,$\pinterjunction$ and $\ptransplication$, as defined in Table~\ref{tab:operators}, are monotonic with respect to $\threeorder$. Following Proposition~\ref{thm:single_monotonicity}, we know that atomic policies are monotonic, and by direct induction, we can conclude that any policy in $\mathcal{P}(\abacm)$ is monotonic, and thus that $\abacm$ is monotonic. 

Now, let $\pi : \abacrequests \to \three$ be a monotonic ideal policy. We show that there exists a policy $P(\pi)$ such that $(\abacm, P(\pi))$ realizes $\pi$.
$(\abacrequests,\abacreqorder)$ is a finite partially ordered set, so we may enumerate its elements using a topological sort.
That is we may write $\abacrequests = [q_0, q_1,\dots,q_n]$, for some $n$ determined by $\mathcal{N}$ and $\mathsf{dom}(\alpha)$, $\alpha \in \mathcal{N}$; and for all $i \leqslant j$, $q_j \not\abacreqorder q_i$.
In particular, we have $q_0 = \emptyset$.

For any non-empty request $q = \set{(\alpha_1,v_1),\dots,(\alpha_m,v_m)}$, we define the policy $t_q = (\alpha_1,v_1) \pand \dots \pand (\alpha_m,v_m)$.
Note that $\semantics{t_q}(q') = \allow$ for all $q' \geqslant q$.
Now, we have $\semantics{t_{q_i}}(q_0) = \na$ for all $q_i$.
Moreover, for $j > i > 0$, we have $\semantics{t_{q_i}}(q_i) = \allow$ and $\semantics{t_{q_j}}(q_i) \ne \allow$.
In other words, for every request $q$ there is a value $m$ such that $\semantics{t_{q_m}}(q) = \allow$ and $\semantics{q_j}(q) \ne \allow$ for all $j > m$.
We now define the policy $\oplus_\pi(t_{q_1},\dots,t_{q_n})$, where
  \[
    \oplus_\pi(d_1,\dots,d_n) =
      \begin{cases}
        \pi(q_m) & \text{where $m = \max\set{i : d_i = \allow}$} \\
        \pi(\emptyset) & \text{otherwise}.
      \end{cases}
  \]
defines an $n$-ary operator $\oplus_\pi$.

We now prove that $\oplus_\pi$ is monotonic. Intuitively, we want to prove that the order over tuples of decisions implies the order over requests, in order to use the monotonicity of $\pi$.
Let $d_1,\dots,d_n \in \three$ and $d'_1,\dots,d'_n \in \three$ with $d'_i \decorder d_i$, $1 \leqslant i \leqslant n$.
Let $q'$ and $q$ be the requests identified by $(d'_1,\dots,d'_n)$ and $(d_1,\dots,d_n)$, respectively.
By definition, we have $\oplus_\pi(d'_1,\dots,d'_n) = \pi(q')$  and $\oplus_\pi(d_1,\dots,d_n) = \pi(q)$.
Furthermore, let ${\sf m}$ be the maximal index such that $d'_{\sf m} = \allow$, it follows that $q' = q_{\sf m}$. Since $d'_{\sf m} \decorder d_{\sf m}$, we can deduce that
$d_{\sf m} = \allow$, implying that $q_{\sf m} \abacreqorder q$, that is, $q' \abacreqorder q$. By hypothesis, $\pi$ is monotonic, and thus we have $\pi(q') \decorder \pi(q)$,
allowing us to conclude that $\oplus_\pi$ is monotonic.

Finally, we know, by a result of Blamey~\cite{blamey02}, that any monotonic operator can be built from $\set{\pnot,\pand,\por,\pinterjunction}$. In other words, the
policy $\oplus_\pi(t_{q_1}, \dots, t_{q_n})$ belongs to $\abacm$, and therefore we can conclude that $\pi$ can be realized by
$(\abacm,\oplus_\pi(t_{q_1}, \dots, t_{q_n}))$.
\end{proof}

% The proof of this result and other results in this section can be found in the appendix.
Theorem~\ref{thm:abac-monotonically-complete} demonstrates that all access control policies built and evaluated using
$\abacm$ are monotonic, and that all monotonic ideal policies can be realized by a policy in $\abacm$. Hence, if policy monotonicity is an important
feature for a system designer, then $\abacm$ provides that guarantee. However, $\abacm$ is not complete, since some (non-monotonic) ideal policies cannot be built from it. We propose in the following section a complete model.

\subsection{$\abacc$: A complete model}
\label{sec:abacc}

In some situations, one might want to define non-monotonic policies or we might want to consider a query set in which the same attribute can have different values within a given query. In such situations, we consider the model $\abacc$ which, in addition
to being complete, is defined over the set of queries $\abacmrequests$.

\begin{Def}
\abacc is defined to be $(\abacmrequests, \mathcal{A}(\mathcal{N}), \three, \set{\pnot, \dbd, \por}, \semantics{\cdot})$. 
\end{Def}

\begin{Thm}\label{thm:abacc-complete}
\abacc is complete.
\end{Thm}
\begin{proof}
The structure of this proof is similar to that of Theorem~\ref{thm:abac-monotonically-complete}, with the difference that the ideal policy $\pi$ need not be monotonic.
We show that there exists a policy $p(\pi)$  in $\abacc$ that realizes $\pi$.

Concretely, we consider the enumeration over the requests, and we build the same operator $\oplus_\pi$. However, in this case, we do not need to prove
the monotonicity of $\oplus_\pi$, and we use instead the fact that the logic  $\set{\allow, \deny, \na, \pnot, \dbd, \por}$ is functionally complete~\cite{post2012},
which ensures that $\oplus_\pi$ can be built from $\set{\pnot, \dbd, \por}$.
\end{proof}

It is trivial to see that by considering $\abacmrequests$, we lose the monotonicity with respect to the inclusion ordering $\abacreqorder$.
In particular, for $v,v' \in \mathsf{dom}(\alpha)$ with $v \ne v'$, we have $\set{(\alpha,v')} \abacreqorder \set{(\alpha,v),(\alpha,v')}$,
but $\deny = \semantics{(\alpha,v)}(\set{(\alpha,v')}) \not\decorder \semantics{(\alpha,v)}(\set{(\alpha,v),(\alpha,v')}) = \allow$.

The function $\semantics{\cdot}$ is monotonic for atomic policies and requests in $\abacmrequests$ if we adopt the ordering $\na < \deny < \allow$.
This means that omitting attributes can cause the evaluation of an atomic policy to change from $\allow$ to $\deny$ or $\na$, or from $\deny$ to $\na$.
While this seems to be reasonable, when combined with operators such as $\pdov$, omitting attributes can cause the evaluation of a policy to change from $\deny$ to $\allow$.
(Thus a user may be able to construct a request $q \subseteq q'$ that is allowed when $q'$ is not.
Any non-monotonic language, such as XACML, incorporates this vulnerability.)
% We established in our work on PTaCL that certain operators (some of which are not used in this paper) are monotonic, while others, such as $\pnot$ and $\por$ are not~\cite{post2012}.

\subsection{ABAC with Explicit Conflict}
\label{app:conflict-abac}

The above choice to evaluate an atomic policy $(\alpha,v)$ to $\allow$ if both $(\alpha,v)$ and $(\alpha,v')$ belongs to the query with $v \ne v'$ and $v,v' \in \mathsf{dom}(\alpha)$ could be regarded as being logically inconsistent in the sense that the request also contains a non-matching value.
One could equally well argue, for example, that the request should evaluate to $\deny$.

In order to cope with such situations, we might, therefore, choose to work with the $4$-valued logic $\four =\set{\allow,\deny,\na,\conflict}$, using $\conflict$ to denote conflicting information (in contrast to $\na$ which signifies lack of information).
We introduce the ordering $\fourorder$, where $d_1 \fourorder d_2$ if, and only if, $d_1 = \bot$, $d_1 = d_2$ or $d_2 = \conflict$, and we define
  \[
    \sem{(\alpha,v)}(q) =
      \begin{cases}
        \conflict & \text{if $q \ni (\alpha,v), (\alpha,v')$ such that $v,v' \in \mathsf{dom}(\alpha)$ and $v' \ne v$}, \\
        \allow & \text{if $q \ni (\alpha,v)$ and $q \not\ni (\alpha,v')$ such that $v' \ne v$},\\
        \deny & \text{if $q \not\ni (\alpha,v)$ and $q \ni (\alpha,v')$ such that $v' \ne v$}, \\
        \na & \text{otherwise}.
      \end{cases}
  \]
The definition of the policy operators $\paov$ and $\pdov$ can be extended to unary operators on $\four$, where
  \[
    \paov d =
      \begin{cases}
        \allow & \text{if $d = \conflict$}, \\
        d & \text{otherwise};
      \end{cases}
    \qquad
    \pdov d =
      \begin{cases}
        \deny & \text{if $d = \conflict$}, \\
        d & \text{otherwise}.
      \end{cases}
  \]
Then the policy $\paov(\alpha,v)$ allows a request $q$ whenever $q$ contains a matching attribute, while the policy $\pdov(\alpha,v)$ denies a request $q$ whenever $q$ contains a non-matching attribute value.
Although the notion of conflicting policy decision has already been studied~\cite{DBLP:journals/tissec/BrunsH11}, to the best of our knowledge, this is the first time this notion of conflict has been used to evaluate targets. Intuitively,
a conflict indicates that the request provides too much information for this particular policy. It is worth observing that atomic policies are monotonic with respect to the ordering $\fourorder$, i.e., for all requests $q$ and $q'$ such that $q \abacmreqorder q'$ and for all atomic policies $(\alpha,v)$, we have:
  \[
    \sem{(\alpha,v)}(q) \fourorder \sem{(\alpha,v)}(q').
  \]

A possible way to extend the operators defined in Table~\ref{tab:operators} is to consider the value $\conflict$ as absorbing: for any operator $\oplus : \three^k \to \three$, we define
the operator $\hat{\oplus} : \four^k \to \four$ as follows:
\[
\hat{\oplus}(d_1, \dots, d_k) =
\begin{cases}
  \conflict & \text{if $d_i = \conflict$ for some $i \in [1,k]$}, \\
  \oplus(d_1, \dots, d_k) & \text{otherwise}.
\end{cases}
\]
Clearly, given an operator $\oplus$ defined over $\three$, if $\oplus$ is monotonic according to $\threeorder$, then $\hat{\oplus}$ is also monotonic with respect to $\fourorder$.
It follows that we can still safely use the operators generated by the operators $\pnot$, $\pand$, $\por$ and $\pinterjunction$, and we can deduce that any realizable policy is also monotonic.
However, we lose the result of monotonic completeness, and we can no longer ensure that any monotonic operator can be generated from this set of operators.
Obtaining such a result requires a deeper study of four-valued logic, and we leave it for future work.

\section{Related Work}

Much of the work on specification of access control languages can be traced back to the early work of Woo and Lam, which considered the possibility that different policy components might evaluate to different authorization decisions~\cite{DBLP:journals/csec/WooL93}.
More recent work has considered larger sets of policy decisions or more complex policy operators (or both), and propose a formal representation of the corresponding metamodel~\cite{DBLP:journals/tissec/BertinoCFP03,bona:alge02,DBLP:journals/tissec/BrunsH11,CH10,cram:nordsec10,post2012,dami:pond01,jajo:flex01,li:acce09,ni:dalg09,xacml2.0,XACML3,wije:prop03}.
The ``metamodels'' in the literature are really attempts to fix an authorization vocabulary, by identifying the sets and relations that will be used to define access control policies.

In contrast, our framework makes very few assumptions about access control models and policies that are written in the context of a model.
In this, our framework most closely resembles the work of Tschantz and Krishnamurthi~\cite{confsacmatTschantzK06}, which considered a number of properties of a policy language, including determinism, totality, safety and monotonicity.
% A policy language is said to be deterministic if the evaluation of any request results in a single decision.
% Our notion of access control model makes any policy total, since a policy is defined to be a (total) function from requests to decisions.
% We might, however, distinguish between an access control model that is conclusive and inconclusive.
% The former would have a decision set equal to $\set{\allow,\deny}$, while an inconclusive model would include $\na$ (and, possibly, additional decisions).
% XACML is neither deterministic nor total (in the sense of~\cite{confsacmatTschantzK06}).
%The gap and conflict analysis of Bruns and Huth in the context of PBel~\cite{DBLP:journals/tissec/BrunsH11} is related to notions of totality and determinism, respectively.

The notion of a monotonic operator (as defined by~\cite{confsacmatTschantzK06}) is somewhat different from ours.
This is in part because a different ordering on the set of decisions $\three$ is used and because monotonicity is concerned with the inclusion of sub-policies and the effect this has on policy evaluation.
This contrasts with our approach, where we are concerned with whether the exclusion of information from a request can influence the decision returned.
(In fact, our concept of monotonicity is closer to the notion of safety defined in~\cite{confsacmatTschantzK06}: if a request $q$ is ``lower'' than $q'$, then the decision returned for $q$ is ``lower'' than that of $q'$.)
We would express their notion of monotonicity in the following way: a policy operator $\oplus$ is monotonic (in the context of model $\mathcal{M}$) if for all $p_1,\dots,p_t \in \mathcal{P}(\mathcal{M})$ and all $q \in \mathcal{Q}$, if $\oplus(p_1\dots,p_t)(q) \in \set{\allow,\deny}$, then $\oplus(p_1,\dots,p_i,p',p_{i+1},p_t)(q) \ne \na$ for any $i$ and any policy $p' \in \mathcal{P}(\mathcal{M})$.
% Thus, our intersection operators are not monotonic in the sense used by Tschantz and Krishnamurthi, since $d \oplus \na = \na$ for any $d$ and any intersection operator $\oplus$.
%In summary, it is possible to define and reason about similar properties within our framework.
Moreover, our framework is concerned with arbitrary authorization vocabularies and queries, unlike that of Tschantz and Krishnamurthi, which focused on the standard subject-object-action request format.
%
% Our framework, however, does not make any assumptions about the structure of requests, something that adds considerable complexity to reasoning about the models that can be defined within the framework.
The only assumption we make is that all policies can be represented using a tree-like structure and that policy decisions can be computed from the values assigned to leaf nodes and the interpretation of the operators at each non-leaf node.
%In short, we attempt---we believe for the first time in the literature---to deconstruct what access control means in an open, attribute-based environment.

In addition, we define the notion of {\em completeness} of a model, which is concerned with the expressivity of the policy operators. %Furthermore, the policy operators used to defined policy terms can be classified according to their properties.
There exists prior work on comparing the expressive power of different access control models or the extent to which one model is able to simulate another~\cite{HaJaMo08,OSM00,TrLi07}.
In this paper, we show how our framework enables us to establish whether a model based on a particular set of atomic policies, decision set and policy connectives is complete.
We can, therefore, compare the completeness of two different models by, for example, fixing an authorization vocabulary and comparing the completeness of models that differ in one or more of the models' components (that is, ones that differ in the set of connectives, decision sets, atomic policies and authorization semantics).
While this is similar in spirit to earlier work, this is not the primary aim of this paper, although it would certainly be a fertile area for future research.

\section{Concluding Remarks}

We have presented a generic framework for specifying access control controls, within which a large variety of access control models arise as special cases, and which allows us to reason about the global properties of such systems.
%Our framework allows us to define concepts of monotonicity and completeness.
%We have shown how these properties depend on the operators used to construct policies and the way in which atomic policies are furnished with semantics.
%We have shown how we can use this framework to define a number of different types of attribute-based access control (ABAC) models, differing in the way in which requests are evaluated with respect to atomic policies.
%
A major strength of our approach is that we do not provide ``yet another access control language''.
The framework is not intended to provide an off-the-shelf policy language or PDP (unlike XACML, for example), nor is it intended to be an access control model (in the style of RBAC96, say).
Rather, we try to model all aspects of an access control system at an abstract level and to provide a framework that can be instantiated in many different ways, depending on the choices made for request attributes, atomic policies, policy decisions and policy evaluation functions.
In doing so we are able to identify
 \begin{inparaenum}[(i)]
  \item how and why an access control system may fail to be sufficiently expressive (completeness), and
  \item how and why having an expressive access control system may lead to vulnerabilities (monotonicity).
 \end{inparaenum}

There are many opportunities for future work.  % , some of which we have discussed in previous sections.
The notions of monotonicity and completeness are examples of general properties of an access control model that we can characterize formally within our framework.
We have already noted that there are at least two alternative semantics for atomic policies having the form $(\alpha,v)$ for a three-valued decision set and even more alternatives for a four-valued decision set.
It would be interesting to see how these alternative semantics affect monotonicity and completeness.
We would like to study the composition of access control models, and under what circumstances composition preserves monotonicity and completeness.
Further properties that are of interest include policy equivalence, policy ordering (where, informally, one policy $P_1$ is ``more restrictive'' than $P_2$ if it denies every request that is denied by $P_2$), which may allow us to define what it means for a realizable policy to be ``optimal'' with respect to an (unrealizable) ideal policy.
Moreover, our definition of monotonicity is dependent on the ordering on the set of decisions.
Monotonicity, in the context of the ordering $\deny < \na < \allow$, for example, is a stronger property than the one we have considered in this paper.
Again, it would be interesting to investigate the appropriateness of different forms of monotonicity. 
Furthermore, although XACML is proven not to be monotonic, it is not known under which conditions it can be monotonically-complete, and if additional operators are needed to prove this property, which is also likely to depend on the decision orderings considered. 
% 
% We also believe our framework could provide a basis for reasoning about the privacy of user attributes.
% It may be, for example, that a privacy-conscious user may wish (or demand) to withhold some attributes from an access control system.
% We conjecture that a property related to monotonicity will enable us to build useful access control systems that enable a user to disclose as little information as is required in order to gain access.
% 
% In addition to the above properties, we believe our framework is expressive enough to characterize a wide range of other properties of interest.
% For instance, the \emph{satisfiability problem} asks whether there exists a valuation on the atomic policies such that the policy is true, in other words whether the policy allows at least one request.
% Again, this is a question of considerable interest in the community~\cite{DBLP:journals/tissec/BrunsH11,fisl:icse05} and one that we believe could be addressed within our framework.
% Such a problem can be extended to \emph{constraint satisfaction problems}, for instance in order to determine which attribute values are missing in a request in order to allow the request.

In this paper, we have assumed that there exists an ideal policy and that such a policy is fixed.
Generally, however, a system evolves over time, and an access control policy will need to be updated to cope with changes to that system that affect the users, resources, or context. 
Thus it may be more realistic to specify an initial ideal policy, which might be extremely simple, and the access control policy that best approximates it, and then define rules by which the access control policy may evolve.
With this in mind, it makes sense to regard the access control policy (or components thereof) as a protected object.
Security is then defined in terms of properties that ``reachable'' access control policies must satisfy.
Typical examples of such properties are ``liveness'' and ``safety''~\cite{Sc00}.
Including administrative policies within our framework and investigating properties such as liveness and safety will be an important aspect of our future work in this area.% 


\begin{thebibliography}{10}
\providecommand{\url}[1]{#1}
\csname url@samestyle\endcsname
\providecommand{\newblock}{\relax}
\providecommand{\bibinfo}[2]{#2}
\providecommand{\BIBentrySTDinterwordspacing}{\spaceskip=0pt\relax}
\providecommand{\BIBentryALTinterwordstretchfactor}{4}
\providecommand{\BIBentryALTinterwordspacing}{\spaceskip=\fontdimen2\font plus
\BIBentryALTinterwordstretchfactor\fontdimen3\font minus
  \fontdimen4\font\relax}
\providecommand{\BIBforeignlanguage}[2]{{%
\expandafter\ifx\csname l@#1\endcsname\relax
\typeout{** WARNING: IEEEtran.bst: No hyphenation pattern has been}%
\typeout{** loaded for the language `#1'. Using the pattern for}%
\typeout{** the default language instead.}%
\else
\language=\csname l@#1\endcsname
\fi
#2}}
\providecommand{\BIBdecl}{\relax}
\BIBdecl

\bibitem{Bishop02}
M.~Bishop, \emph{Computer Security: Art and Science}.\hskip 1em plus 0.5em
  minus 0.4em\relax Addison-Wesley, 2002.

\bibitem{post2012}
J.~Crampton and C.~Morisset, ``{PTaCL}: A language for attribute-based access
  control in open systems,'' in \emph{{Principles of Security and Trust - First
  International Conference, (POST 2012), Proceedings}}, ser. Lecture Notes in
  Computer Science, vol. 7215, 2012, pp. 390--409.

\bibitem{DBLP:conf/esorics/GriesmayerM13}
A.~Griesmayer and C.~Morisset, ``Automated certification of authorisation
  policy resistance,'' in \emph{ESORICS}, ser. Lecture Notes in Computer
  Science, J.~Crampton, S.~Jajodia, and K.~Mayes, Eds., vol. 8134.\hskip 1em
  plus 0.5em minus 0.4em\relax Springer, 2013, pp. 574--591.

\bibitem{Ferraiolo:2008:MMA:1377836.1377860}
D.~Ferraiolo and V.~Atluri, ``A meta model for access control: why is it needed
  and is it even possible to achieve?'' in \emph{Proceedings of the 13th ACM
  symposium on Access control models and technologies}.\hskip 1em plus 0.5em
  minus 0.4em\relax ACM, 2008, pp. 153--154.

\bibitem{XACML3}
\emph{eXtensible Access Control Markup Language (XACML) Version 3.0}, OASIS,
  2010, committee Specification 01.

\bibitem{Ferraiolo92}
D.~F. Ferraiolo and D.~R. Kuhn, ``Role-based access control,'' in
  \emph{Proceedings of the 15th National Computer Security Conference}, 1992,
  pp. 554--563.

\bibitem{Ba09}
S.~Barker, ``The next 700 access control models or a unifying meta-model?'' in
  \emph{SACMAT}, B.~Carminati and J.~Joshi, Eds.\hskip 1em plus 0.5em minus
  0.4em\relax ACM, 2009, pp. 187--196.

\bibitem{harr:prot76}
M.~Harrison, W.~Ruzzo, and J.~Ullman, ``Protection in operating systems,''
  \emph{Communications of the ACM}, vol.~19, no.~8, pp. 461--471, 1976.

\bibitem{Rao:2009:AFI:1542207.1542218}
P.~Rao, D.~Lin, E.~Bertino, N.~Li, and J.~Lobo, ``An algebra for fine-grained
  integration of {XACML} policies,'' in \emph{SACMAT}, B.~Carminati and
  J.~Joshi, Eds.\hskip 1em plus 0.5em minus 0.4em\relax ACM, 2009, pp. 63--72.

\bibitem{DBLP:journals/tissec/BrunsH11}
G.~Bruns and M.~Huth, ``Access control via {Belnap} logic: Intuitive,
  expressive, and analyzable policy composition,'' \emph{ACM Transactions on
  Information and System Security}, vol.~14, no.~1, p.~9, 2011.

\bibitem{confsacmatTschantzK06}
M.~C. Tschantz and S.~Krishnamurthi, ``Towards reasonability properties for
  access-control policy languages,'' in \emph{SACMAT}, D.~F. Ferraiolo and
  I.~Ray, Eds.\hskip 1em plus 0.5em minus 0.4em\relax ACM, 2006, pp. 160--169.

\bibitem{blamey02}
S.~Blamey, \emph{Handbook of Philosophical Logic}.\hskip 1em plus 0.5em minus
  0.4em\relax Kluwer Academic Publishers, 2002, vol.~5, ch. Partial Logic, pp.
  261--353.

\bibitem{CH10}
J.~Crampton and M.~Huth, ``An authorization framework resilient to policy
  evaluation failures,'' in \emph{ESORICS}, ser. Lecture Notes in Computer
  Science, D.~Gritzalis, B.~Preneel, and M.~Theoharidou, Eds., vol. 6345.\hskip
  1em plus 0.5em minus 0.4em\relax Springer, 2010, pp. 472--487.

\bibitem{DBLP:journals/csec/WooL93}
T.~Y.~C. Woo and S.~S. Lam, ``Authorizations in distributed systems: A new
  approach,'' \emph{Journal of Computer Security}, vol.~2, no. 2-3, pp.
  107--136, 1993.

\bibitem{DBLP:journals/tissec/BertinoCFP03}
E.~Bertino, B.~Catania, E.~Ferrari, and P.~Perlasca, ``A logical framework for
  reasoning about access control models,'' \emph{ACM Transactions on
  Information and System Security}, vol.~6, no.~1, pp. 71--127, 2003.

\bibitem{bona:alge02}
P.~Bonatti, S.~{De Capitani Di Vimercati}, and P.~Samarati, ``An algebra for
  composing access control policies,'' \emph{ACM Transactions on Information
  and System Security}, vol.~5, no.~1, pp. 1--35, 2002.

\bibitem{cram:nordsec10}
J.~Crampton and M.~Huth, ``A framework for the modular specification and
  orchestration of authorization policies,'' in \emph{NordSec}, ser. Lecture
  Notes in Computer Science, T.~Aura, K.~J\"arvinen, and K.~Nyberg, Eds., vol.
  7127.\hskip 1em plus 0.5em minus 0.4em\relax Springer, 2010.

\bibitem{dami:pond01}
N.~Damianou, N.~Dulay, E.~Lupu, and M.~Sloman, ``The {Ponder} policy
  specification language,'' in \emph{POLICY}, ser. Lecture Notes in Computer
  Science, M.~Sloman, J.~Lobo, and E.~Lupu, Eds., vol. 1995.\hskip 1em plus
  0.5em minus 0.4em\relax Springer, 2001, pp. 18--38.

\bibitem{jajo:flex01}
S.~Jajodia, P.~Samarati, M.~Sapino, and V.~Subrahmanian, ``Flexible support for
  multiple access control policies,'' \emph{ACM Transactions on Database
  Systems}, vol.~26, no.~2, pp. 214--260, 2001.

\bibitem{li:acce09}
N.~Li, Q.~Wang, W.~H. Qardaji, E.~Bertino, P.~Rao, J.~Lobo, and D.~Lin,
  ``Access control policy combining: theory meets practice,'' in \emph{SACMAT},
  B.~Carminati and J.~Joshi, Eds.\hskip 1em plus 0.5em minus 0.4em\relax ACM,
  2009, pp. 135--144.

\bibitem{ni:dalg09}
Q.~Ni, E.~Bertino, and J.~Lobo, ``D-algebra for composing access control policy
  decisions,'' in \emph{ASIACCS}, W.~Li, W.~Susilo, U.~K. Tupakula,
  R.~Safavi-Naini, and V.~Varadharajan, Eds.\hskip 1em plus 0.5em minus
  0.4em\relax ACM, 2009, pp. 298--309.

\bibitem{xacml2.0}
\emph{eXtensible Access Control Markup Language (XACML) Version 2.0}, OASIS,
  2005, committee Specification.

\bibitem{wije:prop03}
D.~Wijesekera and S.~Jajodia, ``A propositional policy algebra for access
  control,'' \emph{ACM Transactions on Information and System Security},
  vol.~6, no.~2, pp. 286--235, 2003.

\bibitem{HaJaMo08}
L.~Habib, M.~Jaume, and C.~Morisset, ``A formal comparison of the {Bell \&
  LaPadula and RBAC} models,'' in \emph{IAS}, M.~Rak, A.~Abraham, and
  V.~Casola, Eds.\hskip 1em plus 0.5em minus 0.4em\relax IEEE Computer Society,
  2008, pp. 3--8.

\bibitem{OSM00}
S.~Osborn, R.~Sandhu, and Q.~Munawer, ``Configuring role-based access control
  to enforce mandatory and discretionary access control policies,'' \emph{ACM
  Transactions on Information and System Security}, vol.~3, no.~2, pp. 85--106,
  2000.

\bibitem{TrLi07}
M.~V. Tripunitara and N.~Li, ``A theory for comparing the expressive power of
  access control models,'' \emph{Journal of Computer Security}, vol.~15, no.~2,
  pp. 231--272, 2007.

\bibitem{Sc00}
F.~B. Schneider, ``Enforceable security policies,'' \emph{ACM Trans. Inf. Syst.
  Secur.}, vol.~3, no.~1, pp. 30--50, 2000.

\bibitem{sacmat09}
B.~Carminati and J.~Joshi, Eds., \emph{SACMAT 2009, 14th ACM Symposium on
  Access Control Models and Technologies, Stresa, Italy, June 3-5, 2009,
  Proceedings}.\hskip 1em plus 0.5em minus 0.4em\relax ACM, 2009.

\end{thebibliography}
\end{document}